\documentclass[sigconf]{acmart}
\usepackage{dsfont}
\usepackage[utf8]{inputenc}
\usepackage{amsmath}  
\usepackage[mathscr]{eucal}
\usepackage{multirow}
\AtBeginDocument{%
  \providecommand\BibTeX{{%
    \normalfont B\kern-0.5em{\scshape i\kern-0.25em b}\kern-0.8em\TeX}}}

\setcopyright{rightsretained}
\acmConference[SIGIR eCom'21]{ACM SIGIR Workshop on eCommerce}{July 15, 2021}{Virtual Event, Montreal, Canada}
\acmYear{2021}
\copyrightyear{2021}
\makeatletter
\renewcommand\@formatdoi[1]{\ignorespaces}
\makeatother
\acmISBN{}

\begin{document}

\title[``Are You Sure?'']{``Are you sure?'': Preliminary Insights from Scaling Product Comparisons to Multiple Shops}



\author{Patrick John Chia}
\authornote{Main contributor.}
\affiliation{%
  \institution{Coveo}
  \city{Montreal}
  \country{Canada}}
\email{pchia@coveo.com }

\author{Bingqing Yu}
\authornotemark[1]
\affiliation{%
  \institution{Coveo}
  \city{Montreal}
  \country{Canada}}
\email{cyu2@coveo.com}

\author{Jacopo Tagliabue}
\authornote{Corresponding author.}
\affiliation{%
  \institution{Coveo Labs}
  \city{New York}
  \state{NY}
}
\email{jtagliabue@coveo.com}

\renewcommand{\shortauthors}{Chia et al.}

\begin{abstract}
Large eCommerce players introduced comparison tables as a new type of recommendations. However, building comparisons at scale without pre-existing training/taxonomy data remains an open challenge, especially within the operational constraints of shops in the long tail. We present preliminary results from building a comparison pipeline designed to scale in a multi-shop scenario: we describe our design choices and run extensive benchmarks on multiple shops to stress-test it. Finally, we run a small user study on property selection and conclude by discussing potential improvements and highlighting the questions that remain to be addressed. 
\end{abstract}



\maketitle

\section{Introduction}

\begin{figure}
  \centering
  \includegraphics[width=\linewidth]{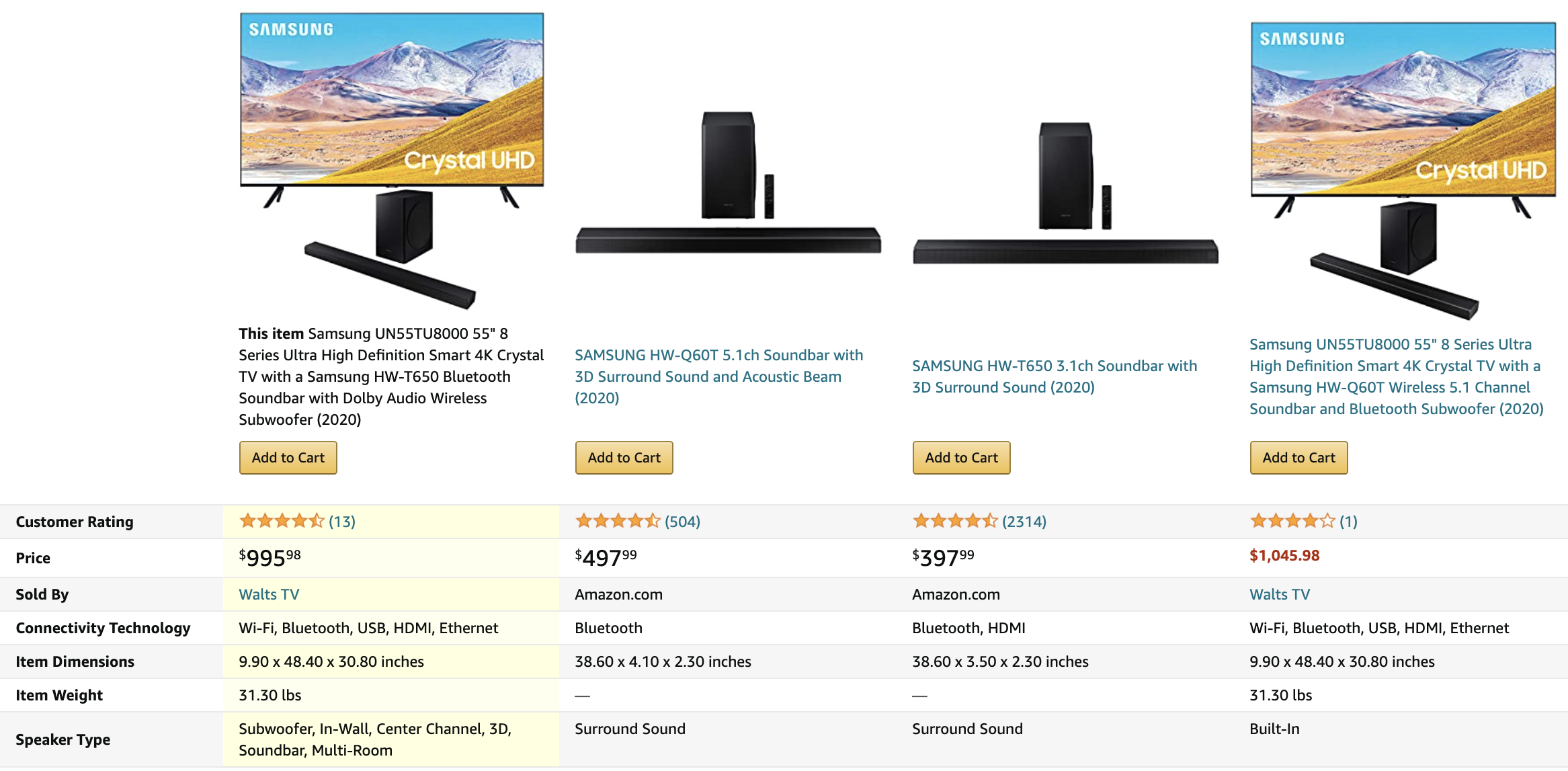}
  \caption{Recommendations as comparisons on \textit{Amazon.com}: the first product (in \textit{yellow}) is the product in the PDP, and the other three are suggestions. As clear from item \#2 and \#3, finding substitutes at scale is all but a trivial task.}
  \Description{Example of product comparisons on \textit{Amazon.com}.}
  \label{pic:amazon_comp}
\end{figure}

Online shopping has seen tremendous growth in recent years~\cite{ecomworld}, and shoppers now face an innumerable number of possibilities, which paradoxically may lead to decreasing satisfaction in their purchase decisions~\cite{10.1086/651235}. Recommender systems (\textbf{RSs}) have been playing an indispensable role in fighting information overload, and large players~\cite{ecomamazon,netflixarticle} have been mostly responsible for modelling and product innovation~\cite{Tsagkias2020ChallengesAR, 10.1145/3219819.3219891}.~\textit{Comparison engines} (\textbf{CEs}) are a special case of RS, in which a product detail page (\textbf{PDP}) displays alternative choices in a table containing informative product specifications (Fig.~\ref{pic:amazon_comp}). Unlike prevalent "More like this" RSs, comparison tables when well-designed not only promote products that are relevant, but also intentionally select products which help customers better understand the \textit{range} of available features. However -- as demonstrated by the sub-optimal alternatives in Fig.~\ref{pic:amazon_comp} -- building a CE is far from trivial even for players with full ownership of the data chain. In~\textit{this} paper, we share preliminary lessons learned when building CEs in a B2B scenario, that is, designing a scalable pipeline that is deployed \textit{across multiple shops}. As convincingly argued in~\cite{10.1145/3383313.3411477,Bianchi2020FantasticEA,CoveoSIGIR2021}, multi-tenant deployments require models to generalize to dozens of different retailers: a successful CE is therefore not only hard to build, but valuable to a wide range of practitioners -- on one side, practitioners outside of humongous websites, who want to enhance their shop in the face of rising pressure from major players; on the other, multi-tenant SaaS providers who need to provide AI-based services that scale to a large number of clients.\footnote{As an indication of the relevant SaaS market size, we witnessed Coveo, Algolia,
Lucidworks and Bloomreach raising more than 100M USD each from venture funds in the last two years for AI-powered services~\cite{CoveoRound,AlgoliaRound,LWRound,BloomreachRound}.} We summarize our contributions as follows:
\begin{itemize}
\item we are the first, to the best of our knowledge, to detail a pipeline for building a comparison engine designed to be scalable in a multi-tenant scenario;
\item we perform extensive experiments on various data cleaning and augmentation approaches. One of our major practical contributions -- in line with what independently reported by \citet{10.1145/3340531.3412732} -- is questioning the widely held belief that co-occurrence patterns are a sufficient proxy for substitutable products~\cite{10.1145/2783258.2783381};
\item we discuss the importance of diversity in the comparison table and propose a decision process to determine relevant attributes. 
\end{itemize}

While we~\textit{do} acknowledge that full online testing is needed to answer some outstanding design questions, we supplement our pipeline tests with a user study, allowing for a preliminary comparison between our results and human judgments, as well as guiding future decisions in our roadmap. Practitioners looking to replicate our work are encouraged to check the \textit{Appendix} for details on our tools, modelling choices and Mechanical Turk setup.


\section{Comparison Engine Pipeline}
\label{sec:pipeline}

In this section, we present the pipeline architecture of our comparison system. The pipeline is composed of three main stages: a first \textbf{candidate fast retrieval} phase, to narrow down the search space; a \textbf{candidate refinement} phase, to ensure precision and produce the final shortlist of products; lastly, a \textbf{final selection} phase, to determine the information to be displayed in the comparison table. We will first explain the logic of each stage, then detail the experiments performed to benchmark the pipeline.\footnote{Note that due to space constraints, we cite the most relevant literature inline at the most appropriate step.}

\begin{figure}
  \centering
  \includegraphics[width=\linewidth]{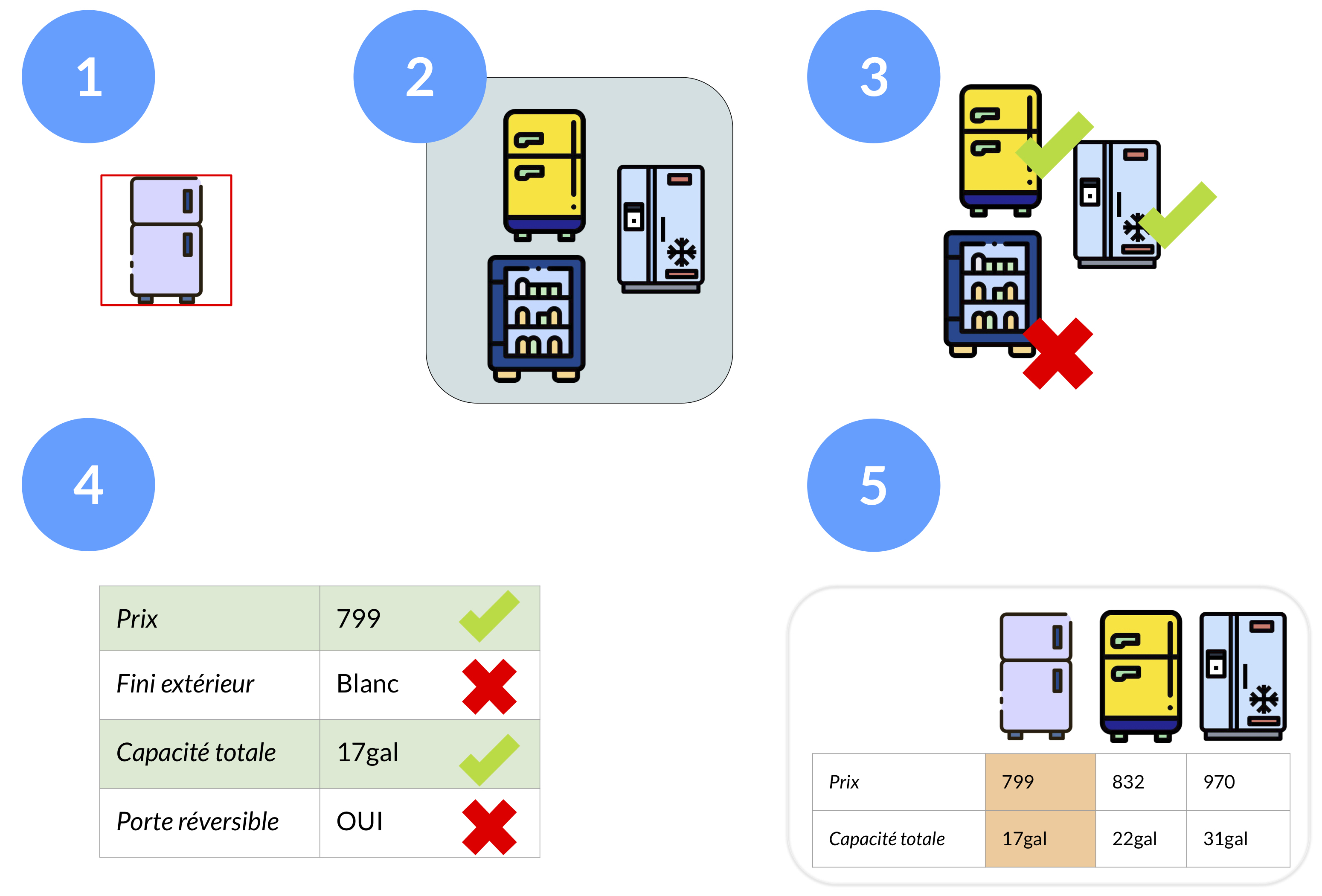}
  \caption{An overview of our CE pipeline: (1) product for the current PDP; (2) fast retrieval of candidate substitutes, focusing on recall; (3) refinement of candidate substitutes using binary classification model; (4) selection of important properties; (5) final selection of substitutes to display in comparison table.
  }
  \Description{Overview of our CE pipeline.}
  \label{pic:pipelinepic}
  \vspace{-4mm}
\end{figure}

\begin{table*}
\centering
\begin{tabular}{c|ccc|ccc}  \toprule
\multicolumn{1}{c|}{} &  \multicolumn{3}{c|}{Shop A} & \multicolumn{3}{|c}{Shop B} \\ \midrule
   Configuration      & P@R=0.7 & P@R=0.8 &P@R=0.9              & P@R=0.7         & P@R=0.8        & P@R=0.9         \\ \midrule
   \textit{Baseline}  & 0.744            & 0.743            & 0.682             & 0.611           & 0.573          & 0.539          \\
   \textit{C=0; S=0}  & 0.759 (0.0235)   & 0.710(0.0212)    & 0.645 (0.0195)    & 0.734 (0.0324)  & 0.690 (0.0333) & 0.632 (0.0290) \\
   \textit{C=0; S=1}  & 0.766 (0.0257)   & 0.723 (0.0237)   & 0.659 (0.0189)    & 0.755 (0.0379)  & 0.706 (0.0423) & 0.643 (0.0382) \\
   \textit{C=1; S=0}  & 0.833 (0.0162)   & 0.802 (0.0226)   & 0.740 (0.0301)    & 0.777 (0.0162)  & 0.732 (0.0131) & 0.658 (0.0123) \\
   \textit{C=1; S=1}  & \textbf{0.842} (0.0150)   & \textbf{0.812} (0.0208)   & \textbf{0.753} (0.0280)  & \textbf{0.789} (0.0189)  & \textbf{0.743} (0.0196) & \textbf{0.663} (0.0176)     \\
   \bottomrule
\end{tabular}
\caption{Precision@Recall = \{0.7,0.8,0.9\} for various configurations for Shop A and B; P@R=X denotes Precision at Recall of X.}
\label{tab:substitute_A}
\vspace{-7mm}
\end{table*}

\subsection{Fast Retrieval}
\label{sec:fastRetrieval}

Candidate retrieval aims to quickly generate potential substitutes given a query product, with a focus on \textit{recall} (step $2$ in Fig.~\ref{pic:pipelinepic}): we try to get a more diverse set, knowing false positives will be screened out in later phases. A common practice for fast retrieval in a dense space is using k-NN over an embedding space~\cite{Zhao2018LearningIE,45530}: since recent literature~\cite{Bianchi2020FantasticEA,coveoNAACLINDUSTRY21,CoveoECNLP202} provides extensive evidence on the representational qualities of behavioral embeddings, we train a \textit{prod2vec} space~\cite{Grbovic15} by adapting word2vec~\cite{Mikolov2013EfficientEO} to eCommerce -- i.e. a \textit{prod2vec} space is just a word2vec space, where words in a sentence are replaced with products in a shopping session (Appendix \ref{appendix:unsupervised}). After obtaining a \textit{prod2vec} space we apply k-NN (based on cosine distance) to retrieve the closest $k=100$ products as its substitute candidates. Analogous to words in word2vec, products which are distributionally similar (based on historical sessions) are close in the \textit{prod2vec} space, therefore the candidates retrieved in this step are already biased towards substitutable products.

\subsection{Candidate Refinement}
\label{sec:refinement}

Candidates produced by the first stage are passed to the second stage for fine-grained processing. The goal of this stage is to boost precision by filtering out candidates that do not have matching product type, and re-rank the remaining ones so the most comparable ones are at the top of the list (step $3$ in Fig.~\ref{pic:pipelinepic}). We employ a binary classification model (i.e. given a pair of products, are they substitutes?) built on top of a Siamese Network \cite{journals/ijprai/BromleyBBGLMSS93}, fed with unsupervised behavioral data.

\subsubsection{Unsupervised Behavioural Data}

Generating training data for substitute product detection is a well-explored topic in the literature~\cite{10.1145/2783258.2783381,Chen2020TryTI,ijcai2019-598}. However, our inference is somewhat harder than a general substitute classifier where products are sampled from the entire catalog, as our model needs to be able to make subtler distinctions among a selected group of candidates that have been shortlisted by a coarse similarity measure (Section~\ref{sec:fastRetrieval}). To overcome the problem of naive sampling and reduce the noise in behavioral data, we built a three-step process generating the final training set, with free parameters, $M$, $N$, $Z$ (Appendix~\ref{appendix:unsupervised}):

\begin{enumerate}
    \item We use \textbf{co-view} and \textbf{co-purchase} patterns to obtain positive and negative training examples. Positive examples are obtained from pairs of products which are viewed consecutively (co-view: if I want a TV, I will check several TVs in a row) and negative examples are obtained from products which are purchased consecutively (co-purchase: if I just bought a TV, I am unlikely to buy a second one). To reduce noise, we set a minimum threshold for the number of co-view occurrences ($N$) and the number of co-purchase occurrences ($M$) for a pair to be considered a positive or negative example respectively.
    \item We intuit that substitutable products are \textit{a priori} visually similar, and utilize this to further reduce noise in the data. Thus, we apply a threshold on the cosine similarity of the image embedding of pairs to further refine this set of training examples. Given an image vector obtained through a pre-trained VGG16~\cite{Simonyan15}, we enforce that positive/negative pairs must have a minimum/maximum cosine similarity.\footnote{While drafting this paper, we realized a similar approach has been recommended independently by~\cite{kdd_amazon}.}  We refer to this refinement/cleaning process as \textbf{C}.
    \item We remove pairs which are given both positive and negative labels, then build a graph using the remaining positive pairs and extract disconnected subgraphs as clusters of substitutable products. We eliminate clusters of size $>Z$ when generating synthetic pairs, to reduce the risk of sampling from clusters formed by noisy pairs and, at the same time, improve the balance of product types in the training data. By taking an existing positive/negative pair from our behavioural logs, we generate synthetic pairs by swapping out one of the products in the original pair with any product found in its substitute cluster, unlike \cite{guo-etal-2020-deep} which samples negative examples from a random disconnected subgraph. We refer to this augmentation process as \textbf{S}.
\end{enumerate}

We emphasize that \textit{only} behavioral logs and product images are necessary so far: our approach does not assume peculiar meta-data or pre-made taxonomy, nor does our classifier require costly labelling, making the pipeline suitable for multi-shop scaling.

\subsubsection{Binary Classifier: A Siamese Network}
We utilize a binary classifier to predict whether two input products are substitutes or not. Products are represented by various dense representations of product features, such as behavioural embeddings and word2vec embeddings for product title, description and category strings (See Appendix \ref{appendix:unsupervised}). For full reproducibility, we provide architectural and hyper-parameter details in Appendices \ref{appendix:unsupervised} \& \ref{appendix:model}.

\subsection{Product and Property Selection}
\subsubsection{Relevant Property Selection}

At this stage, we make the only significant meta-data assumption of the entire pipeline, that is, the target catalog should specify product properties in \textit{some} structured way -- based on our experience with dozens of deployments, this is not a universal feature, but it is  common for verticals with technical products (DIY, electronics, etc.), for which CEs are most useful. Given a mapping from products to their properties (say, from TV $123$ to the set <\textit{resolution}, \textit{screen size}, ...>), this stage determines which properties are \textit{relevant} to shoppers when they are making a purchase decision (step $4$ in Fig.~\ref{pic:pipelinepic}). By passing the candidates from Section~\ref{sec:fastRetrieval} to the classifier in~\ref{sec:refinement}, we generate a final list of substitutable products, given an initial query product. For this list, we rank properties $P_1, P_2, ..., P_n$ based on the weighted sum of three components\footnote{Weights have been determined empirically at first, but see Section~\ref{sec:mturk} and our conclusion for potential use of human-in-the-loop inference.}, highlighted as important by previous literature \cite{inproceedingsattrank,Dong2020AutoKnowSK} and domain knowledge:  

\begin{enumerate}
    \item \textit{Query frequency}: properties which are important to shoppers tend to appear frequently in shopper-generated content \cite{10.1145/2857054,10.1145/3343413.3377956} such as queries \cite{inproceedingsattrank}. We calculate the query frequency for each property (and their possible values) by mining search logs, and normalize the counts to range $[0,1]$;
    \item \textit{PDP frequency}: merchandisers are more likely to explicitly mention important attributes in the PDP. We calculate a normalized count for each attribute by mining product descriptions in the catalog;
    \item \textit{Property entropy}: it is important, for meaningful comparison, that property values have enough variation, so that comparison tables can help navigate easily the possible dimensions of a catalog. To calculate variety, we measure the entropy of the distribution of property values across the list of substitute products.
\end{enumerate}

\subsubsection{Final Display Selection}
Recent literature \cite{wu2019recent} has highlighted the importance of diversity in RSs. Thus, after determining important product properties, we select the final $W=3$ substitutes per query item (step $5$ in Fig.~\ref{pic:pipelinepic}), by making two additional calculations: \textit{price diversification} and \textit{representative selection}. Given the list of substitutes, we group products into 7 bins based on their log price.\footnote{The mean log-price is used to set the central bin and the standard deviation is used to determine the bin width.} We discard the first and the last bin, as extreme prices can signal a potential mismatch in product category. With 5 bins remaining, we sample one substitute from the same price bin as the query item, and two substitutes from its higher-pricing and lower-pricing neighbor bin. Finally, similarly to \cite{10.1145/1148170.1148245}, we employ a greedy approach during sampling, which maximises the information diversity among the final products to be displayed. We represent the property values of each product via one-hot encoding, so that products are represented by a concatenation of their one-hot encoded property vectors. We compute the difference in their information content by Hamming distance, where each property is weighted by the negative exponential of the entropy of the distribution of the property's values. The intuition is that we want to vary properties which are far from quasi-uniform distributions to display products with meaningful variation, thereby giving shoppers a more complete picture of what is available.

\section{Experiments}

After having discussed the pipeline design, we report our experiments for the substitute model and the user study performed on property selection.

\subsection{Substitute model}

We evaluate the effectiveness of training a neural model for substitute classification in an unsupervised manner, by leveraging a manually prepared held out set for benchmarks (Section~\ref{sec:dataset})\footnote{Since in a real deployment labels will not be present, a research setting is needed to first validate how well unsupervised training performs on golden data.}. Since the objective of the substitute model is to refine the candidates from the initial fast retrieval step (Section \ref{sec:fastRetrieval}), where candidates are \textit{a priori} likely, but not guaranteed, to be substitutes, our test set also mimics this distribution. As a baseline, we thus adopt the cosine similarity (re-scaled to $[0,1]$) between the image vectors of two products as the confidence score for substitutability. This serves as a simple yet realistic baseline that allows us to quantitatively assess the precision boost afforded by the substitute model.

For our experiments, we consider all configurations of image vectors for cleaning ($\textbf{C}\in \{0,1\}$) and synthetic augmentation ($\textbf{S}\in \{0,1\}$) to shed light on their contribution to performance.\footnote{1 denotes usage/application of method whereas 0 denotes non-usage.} We run experiments on 3 different seeds and with various combinations of dense product representation (Appendix \ref{appendix:unsupervised} \& \ref{appendix:model}) as input. For each configuration of \textbf{C} and \textbf{S}, we report average performance across seeds and product representations used, as well the confidence intervals in plots. We run an extensive set of experiments to acknowledge the varying quality of such representations across catalogs, and to demonstrate robustness of certain configurations when scaling CEs in a multi-shop scenario.

\subsubsection{Dataset}
\label{sec:dataset}
For training and validation, we extract unsupervised co-view and co-purchase data from shopping sessions of two partnering shops, \textbf{Shop A} and \textbf{Shop B}. They are mid-sized shops: Shop A is in the sport apparel industry whereas Shop B is in home improvement. We use $80\%$ of all products for training and the remaining $20\%$ for validation. We consider this to be a strict testing regime as none of the products used in validation and testing are seen in training. For testing, we first obtain a golden mapping of clusters of substitutable products by heuristic matching of categories provided in catalog data and extensive manual filtering. The golden mapping is then used to generate positive and negative test examples as explained in Section~\ref{sec:fastRetrieval}.\footnote{Full descriptive statistics are reported in Appendix \ref{appendix:unsupervised}.} We selected shops with catalogs that are of high quality and contain fine-grained category information in order to generate golden mappings which best capture product substitutability. We emphasize that such catalog quality is not guaranteed across shops, which motivates our use of unsupervised data.  

\subsubsection{Results}

\begin{figure}
  \centering
  \includegraphics[scale=0.45]{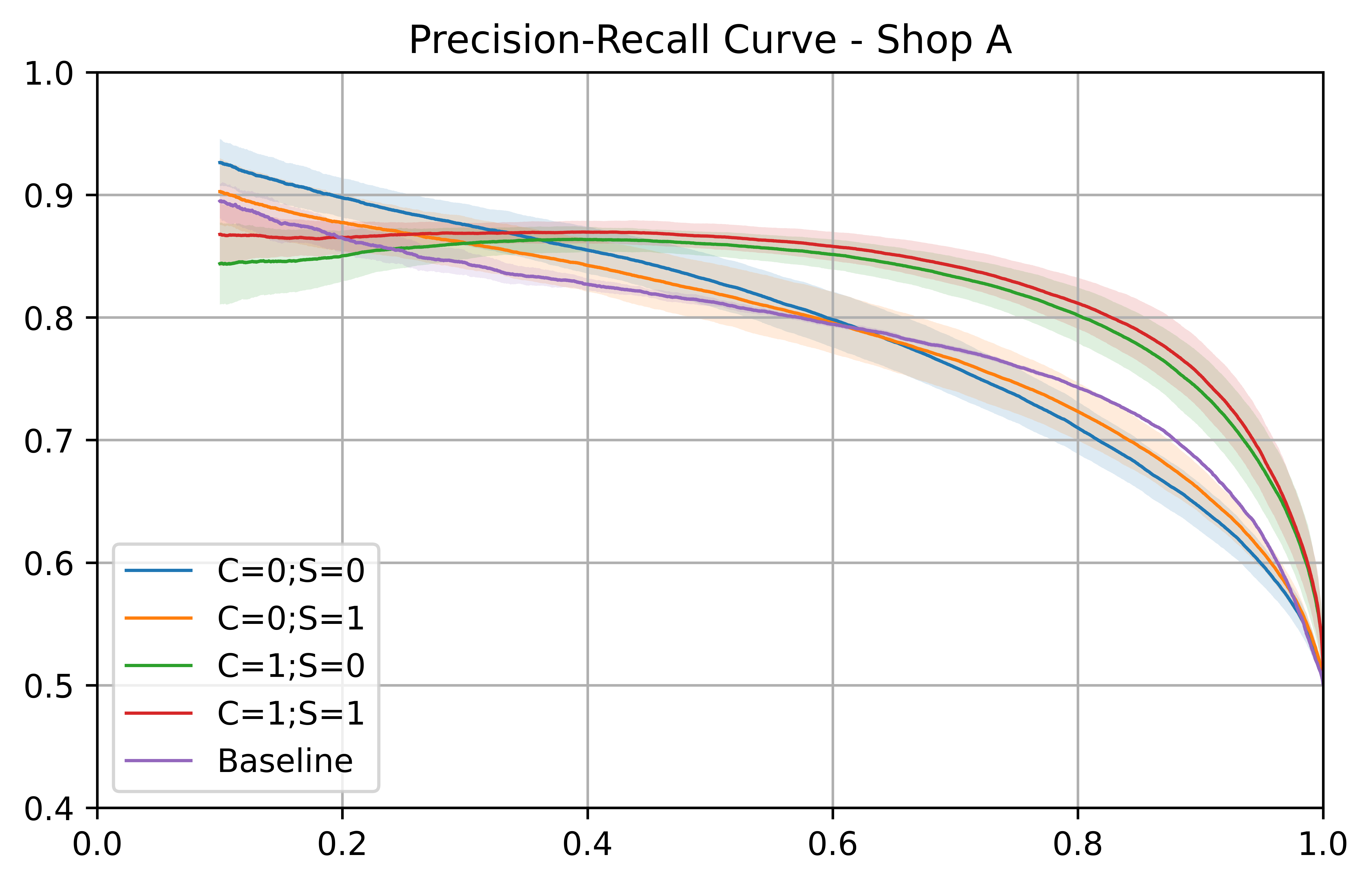}
  \includegraphics[scale=0.45]{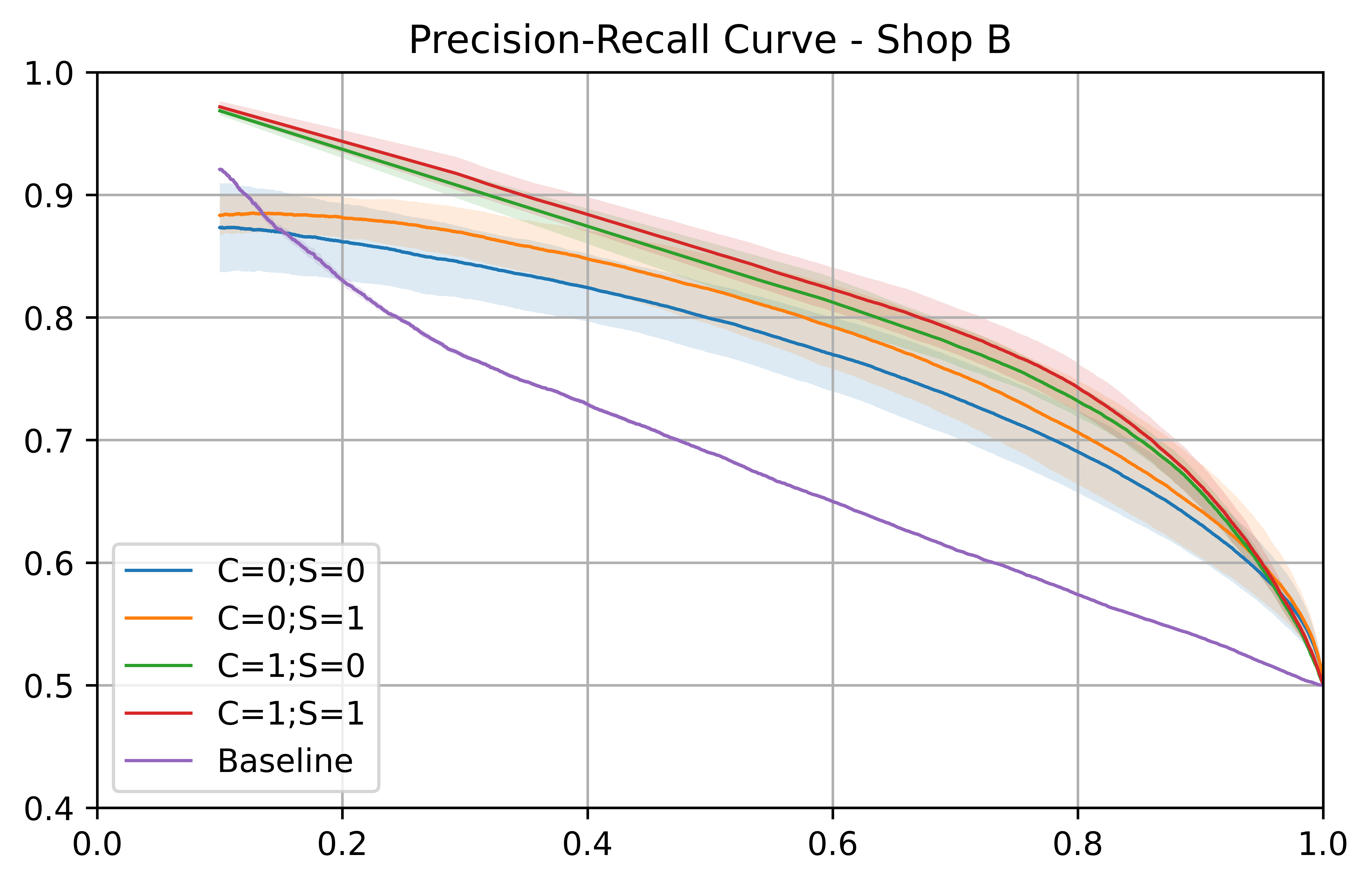}
  \caption{PR Curve for various configurations of C and S, and the Baseline for Shop A and B. Results are the average across seeds and input features, with a confidence interval of +/- 1 SD.}
  \label{pic:pr_curves}
  \vspace{-5mm}
\end{figure}

We summarize experimental results in Table \ref{tab:substitute_A}, and plot in Fig. \ref{pic:pr_curves} the Precision-Recall (PR) curves. For Shop A, when image vectors are not used for cleaning ($C=0$), the model performs only as good as the baseline. When $C=1$, we see a significant increase in precision across the higher ranges of recall; on the other hand, synthetic augmentation, $S$, has minimal effect on model performance. Similar trends are observed for Shop B, albeit the benefit of $C=1$ is less pronounced. These results demonstrate the effectiveness of using image vectors to clean the otherwise noisy unsupervised co-occurrence data, and validate the effectiveness of the preparation detailed in Section~\ref{sec:refinement}. However, as evident in the baseline performance of Shop B, caution must still be taken when relying on image vectors -- depending on the vertical, visual similarity may not be as strong a proxy for substitutability and/or the pre-trained models used to generate the image embeddings are not fine-tuned for products in certain verticals. This opens up interesting avenues for future work such as self-supervised learning \cite{DBLP:journals/corr/abs-2103-03230} for niche verticals.

\subsection{Property selection}
\label{sec:mturk}
We run an Amazon Mechanical Turk (MTurk) study to get preliminary insights on how well our algorithm matches how shoppers rank properties. Our investigation involves 4 product types that range from known products (e.g. running shoes) to increasingly technical (e.g. ski), each with 5-8 properties. While agreement with human judgement varies depending on the category, the algorithm seems to pick up at least \textit{some} qualitatively relevant latent dimension.

\subsubsection{Data Collection}
\begin{figure}
  \centering
  \includegraphics[scale=0.32]{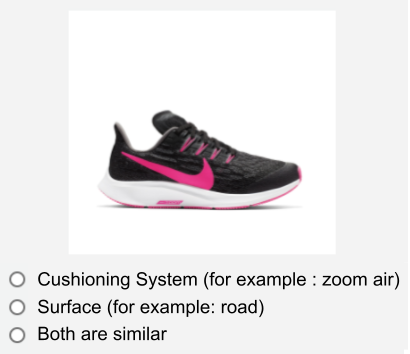}
  \caption{Example of a HIT task.}
  \label{pic:prop_question}
  \vspace{-6mm}
\end{figure}

We collect pairwise human judgements on property preferences. For each comparison, we present workers with the image of a product and two of its properties and ask them to judge which is more important to them when making a purchase decision. Each Human Intelligence Task (HIT) has 3 comparisons (Fig. \ref{pic:prop_question}) in addition to a control task to filter out low quality responses. We collected an average of 30 responses per property pair for this experiment. To collate pairwise human responses, we estimate the underlying ranking using the Bradley-Terry Model \cite{10.2307/2334029, choix}. We compare the estimated ranked list against our algorithm using Rank-biased Overlap \cite{10.1145/1852102.1852106} (RBO) as the measure of agreement.

\subsubsection{Results}
The results are summarized in Table \ref{tab:property}. Agreement between our algorithm and humans is higher for popular/common products, lower for highly-technical ones, which may also reflect a lack of domain-specific knowledge by general MTurk workers.\footnote{Anecdotally, we also solicited feedback from active skiers in \textit{Coveo}, and found that their experience influenced the properties which they found important.} Interestingly enough, the RBO for \textit{Running Shoes} is by far the highest. We suspect that this is because \textit{Running Shoes} lie at an intersection of being both well-known, whereby crowded-sourced responses are most reliable, and technical, such that there exists a stronger ranking/ordering of its properties.

\begin{table}[h]
\centering
\begin{tabular}{lc}   \toprule
   Product          & RBO        \\ \midrule
   Shirt            & 0.633      \\
   Shorts           & 0.483      \\
   Running Shoe     & \textbf{0.783}      \\
   Ski              & 0.169      \\
   \bottomrule
\end{tabular}
\caption{RBO for human vs our ranking (best in bold); $RBO=0.6$ for random permutation of length 5.}
\label{tab:property}
\vspace{-8mm}
\end{table}

\section{Conclusion}
We shared insights from building a CE addressing large-to-mid-shops in the market long-tail, and as such particularly suited for multi-shop deployment. While preliminary, our multi-shop benchmarks confirms the viability of our pipeline, and we look forward to testing it online. Two important areas of improvements are \textit{personalization} and \textit{human-in-the-loop} inference. In the current system, all shoppers would receive the same set of candidates, but individual preferences and session intent~\cite{CoveoECNLP202} may be used to further shape the final table. 

Finally, of the three ways in which we could use human judgements -- qualitative validation, training data and active learning -- we just focused on the first. Given the scalability of MTurk, however, we plan on extending human-in-the-loop computation in further iterations of the project.

\section{Ethical Considerations}
User data has been collected in the process of providing business services to the clients of~\textit{Coveo}: user data is collected and processed in an anonymized fashion, in full compliance with existing legislation (GDPR). In particular, the target dataset uses only anonymous uuids to label sessions
and, as such, it does not contain any information that can be linked to individuals. As explained, our MTurk HITs include a task with pre-defined answer to control for workers randomly answering to questions; however, we still compensate workers for their time, even if their answers get discarded from the analysis.

\begin{acks}
We wish to thank Federico Bianchi, Mattia Pavoni and Andrea Polonioli for comments on a previous draft of this work, and general support with this research project.
\end{acks}

\bibliographystyle{ACM-Reference-Format}
\bibliography{short_refs}

\appendix

\section{Implementation Details}

We implement our pipeline leveraging \textit{Metaflow}~\cite{Metaflow}, which allows us to programmatically define our pipeline as a DAG. We develop our pipeline with three core phases (spread across several steps): 

\begin{itemize}
    \item we dedicate initial steps in the DAG to pull data (such as user sessions, pre-cached embeddings) from various sources like Snowflake and S3, and perform various transformations on the data. Note that many of these steps run in parallel;
    \item we launch in parallel our model training (which may in itself contain several steps as outlined in Section \ref{sec:pipeline}) with various configurations (e.g. input features). In addition, we are able to dedicate steps which have high resource demands (e.g. GPU) to AWS Batch;
    \item we collate the results (e.g. metrics, trained model, model predictions) from each parallel run, and store them as Data Artifacts on S3 for further analysis.
\end{itemize}

\begin{figure}[h]
  \centering
  \includegraphics[scale=0.2]{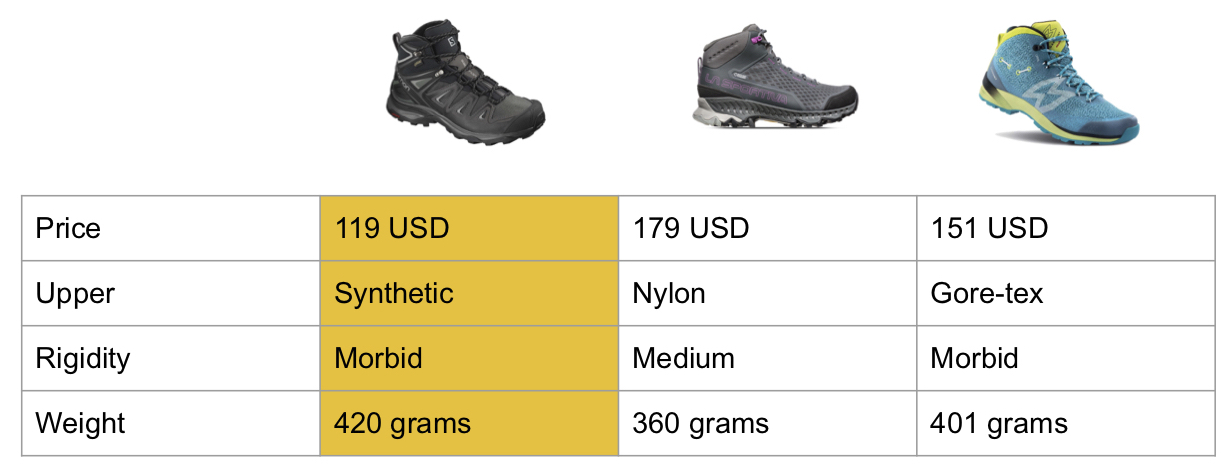}
  \caption{Example of a comparison table for mountain shoes, as prepared by our pipeline.}
  \label{pic:table}
  \vspace{-3mm}
\end{figure}

The adoption of \textit{Metaflow} on top of our cloud provider (\textit{AWS}) speeds up development time (since it is the same code running locally and remotely), reduces training time (thanks to parallelism and GPU provisioning) and increases confidence in our experiments (thanks to versioning and full pipeline replayability). The setup we adopt fully decouples writing code from the underlying infrastructure, including data retrieval thanks to the ``PaaS-like feeling'' of Snowflake~\cite{10.1145/2882903.2903741}. Fig.~\ref{pic:table} shows the comparison table for a pair of mountain shoes (yellow), as produced by our Metaflow pipeline.

\section{Unsupervised Data and Product Representations}
\label{appendix:unsupervised}
In this section, we provide details and hyper-parameters used in the generation of training data and of dense unsupervised representations for products.

\subsection{Data Preparation}
\begin{itemize}
    \item Co-view and Co-purchase Data: For Shop A, we obtain shopping sessions over a period of 3 months and for Shop B we obtain shopping sessions over a period of 1 month. For co-view pairs, we enforce a minimum count, $N=10$, and for co-purchase pairs we enforce a minimum count $M=1$.
    
    \item Cleaning with Image Vectors: For both Shop A and Shop B, we enforce that positive pairs have a cosine similarity $\ge0.8$ and that negative pairs have a cosine similarity $\le0.5$.

    \item Synthetic Augmentation: Maximum cluster size, $Z$, is set to 40.
\end{itemize}

Refer to Tables \ref{tab:co_stats} \& \ref{tab:data_stats} for descriptive statistics of session and training data.

\begin{table}[H]
\centering
\begin{tabular}{cccc}  \toprule
   Shop    & \#Products & \# Browse Session  & \# Purchase Session \\ \midrule
   Shop A  & 20k        & 1.5M               & 27k                 \\
   Shop B  & 50k        & 3M                 & 12K                 \\
   \bottomrule
\end{tabular}
\caption{Descriptive statistics of session data.}
\label{tab:co_stats}
\vspace{-8mm}
\end{table}

\begin{table}[H]
\centering
\begin{tabular}{cccc}  \toprule
    Shop                     & Config            & Train (Pos/Neg)  & Validation (Pos/Neg)   \\ \midrule
    \multirow{4}{4em}{Shop A}& \textit{C=0; S=0} & 19k/18k          & 1.5k/1k   \\
                             & \textit{C=0; S=1} & 27k/75k          & 2.8k/3.7k \\
                             & \textit{C=1; S=0} & 8k/6k            & 0.5k/0.5k \\
                             & \textit{C=1; S=1} & 17k/33k          & 1.5k/1.5k \\ \midrule
    \multirow{4}{4em}{Shop B}& \textit{C=0; S=0} & 50k/20k          & 3k/1k     \\
                             & \textit{C=0; S=1} & 60k/40k          & 6k/5k     \\
                             & \textit{C=1; S=0} & 40k/10k          & 2.5k/1k   \\
                             & \textit{C=1; S=1} & 70k/120k         & 7k/7k     \\            
            
   \bottomrule
\end{tabular}
\caption{Descriptive statistics of training data.}
\label{tab:data_stats}
\vspace{-8mm}
\end{table}

\subsection{Unsupervised Product Representations}
\begin{itemize}
    \item Prod2Vec Embeddings: We train behavioural product embeddings using CBOW with negative sampling \cite{Mikolov2013EfficientEO}, swapping the concept of words in a sentence with
products in a browsing session. Following best practices of \cite{Bianchi2020FantasticEA} we adopt the hyper-parameters: \textit{window = 5 , iterations = 30, ns\_exponent = 0.75, dimensions = 48}, with the exception of a smaller window size, so that more emphasis is placed on co-viewed, and hence more likely substitutable products.
    
    \item Textual Embeddings: We train Textual Embeddings using CBOW with negative sampling and using product descriptions as our text corpus. We adopt the hyper-parameters: \textit{window = 10, iterations = 30, ns\_exponent = 0.75, dimensions = 48}. We then take the name, description and categories of each product and obtain a dense representation for each meta-data by applying average-pooling over their word representations.
    
    \item Image Embeddings: We prepare Image Embeddings by utilising a pre-trained VGG16 \cite{Simonyan15} network, and apply 7x7 2D-MaxPooling to the final MaxPool layer of VGG16 to obtain a 512-dim representation. 
\end{itemize}

\section{Model Architecture and Training}
\label{appendix:model}
\subsection{Model Architecture}
In this section we provide architectural details on the binary comparison model. At a high level, the model takes in two products as inputs and provides a confidence score indicating of whether the two products are substitutes. 

First, each product $p_i$ is represented by $k$ embeddings $[ f_i^0,..., f_i^k ]$, each of dimension $D$ representing a different type of information or modality. Details on how these embeddings are obtained can be found in Appendix \ref{appendix:unsupervised}. 

Secondly, the embeddings of a product are fused into a single dense representation by a neural network $FUSE(f^0,...,f^k)$, which is re-used across all products. We define $FUSE(f^0,...,f^k)$ as:

\begin{equation*}
    FUSE(f^0, ..., f^k) = N(\rho( C( [ \phi^0(f^0), ..., \phi^k(f^k) ] ) ))
\end{equation*}

where $\phi^k$ is a dense re-projection layer (48-dim, ReLU activation), $C$ is the concatenation operation, $\rho$ is a dense fusion layer (128-dim, ReLU activation) and $N$ refers to L2-Normalization operator. 

Lastly, the fused representations of two products, $h_1, h_2$ are passed into a neural network $BIN(h_1,h_2)$, which produces the confidence score. We define $BIN(h_1,h_2)$ as:

\begin{equation*}
    BIN(h_1,h_2) =  \sigma(\psi(|h_1 - h_2|))
\end{equation*}

That is, we take the element-wise absolute difference \cite{reimers2019sentencebert} between the two inputs and pass it into a dense classification layer $\psi$ (1-dim) followed by the sigmoid function $\sigma$ to produce the binary classification score.

\subsection{Model Training}
For all experiments, we use Adam optimizer with learning rate of $0.001$, early stopping with patience of 20 epochs and a batch size of 32. For all experiments we tested the follow configurations of product representations:
\begin{itemize}
    \item description, name, prod2vec;
    \item categories, description, name;
    \item categories, description, name, prod2vec.
\end{itemize}

The feature set that yielded best results is [categories, description, name, prod2vec].
      
\end{document}